\documentclass[aps,pra,reprint]{revtex4}
\usepackage{graphicx}
\begin{document}
\title{New application of decomposition of U(1) gauge potential:\\
        Aharonov-Bohm effect and Anderson-Higgs mechanism}
\author{Jian-Feng Li$^{1}$, Yu Jiang$^{2}$, Wei-Min Sun$^{3,4}$,
Hong-Shi Zong$^{3,4}$ and Fan Wang$^{3,4}$}
\address{$^{1}$ College of  Mathematics and Physics, Nantong University, Nantong 226019, China}
\address{$^{2}$ College of Mathematics, Physics and Information Engineering, Zhejiang Normal University, Jinhua 321004, China}
\address{$^{3}$ Department of Physics, Nanjing University, Nanjing 210093, China}
\address{$^{4}$ Joint Center for Particle, Nuclear Physics and Cosmology, Nanjing 210093, China}

\begin{abstract}
In this paper we study the Aharonov-Bohm (A-B) effect and
Anderson-Higgs mechanism in Ginzburg-Landau model of superconductors
from the perspective of the decomposition of U(1) gauge potential.
By the Helmholtz theorem, we derive exactly the expression
of the transverse gauge potential $\vec{A}_\perp$ in A-B experiment,
which is gauge-invariant and physical. For the case of a bulk
superconductor, we find that the gradient of the total phase field
$\theta$ provides the longitudinal component ${\vec A}_{\parallel}$,
which reflects the Anderson-Higgs mechanism. For the case of a
superconductor ring, the gradient of the longitudinal phase field
$\theta_1$ provides the longitudinal component ${\vec
A}_{\parallel}$, while the transverse phase field $\theta_2$
produces new physical effects such as the flux quantization inside a
superconducting ring.

\bigskip
Key-words:  gauge potential decomposition; pure gauge; phase field; superconducting ring

\bigskip
E-mail: sunwm@chenwang.nju.edu.cn. ~~~~PACS Numbers: 03.65.Vf, 71.35.Lk
\end{abstract}

\maketitle
The decomposition of the gauge potential in gauge field theory plays an important role in the study of many physical problems. In the literature there are different types of gauge potential decomposition in different physical contexts \cite{Duan,Cho,Faddeev}. For the case of U(1) gauge theory, the one with the most definite physical meaning is the decomposition of the vector gauge potential ${\vec A}$
into its transverse component $\vec{A}_\perp$ and longitudinal component $\vec{A}_\parallel$. This type of decomposition has many applications, e.g., it can be used to construct a decomposition of the QED angular momentum into the spin and orbital parts of the electron and the photon
with each part satisfying the angular momentum algebra and the requirement of gauge invariance simultaneouly \cite{a4}, and the proper gauge invariant momentum and Hamiltonian operator for a charged particle in an external
electromagnetic field \cite{Wang,Sun}. In this paper we shall discuss two physical problems, the Aharonov-Bohm (A-B) effect \cite{a1} and the the Anderson-Higgs (A-H) mechanism \cite{a2} in Ginzburg-Landau (G-L) model \cite{a3} for the superconductors, from the perspective of this type of decomposition.

To start, we recall the decomposition of the vector potential ${\vec A}$ in terms of
its transverse and longitudinal parts
\begin{equation}\label{decomposition}
 \vec A=\vec A_{\bot}+\vec A_{\parallel},
\end{equation}
where $\vec A_{\bot}$ and $\vec A_{\parallel}$ are defined by
\begin{eqnarray}
\vec{\nabla} \cdot \vec{A}_{\bot}&=& 0 \nonumber \\
\vec{\nabla} \times \vec A_{\parallel} &=& 0.
\end{eqnarray}
With the boundary condition that $\vec A$, $\vec A_{\bot}$ and $\vec A_{\parallel}$ all vanish at spatial infinity, the above two conditions prescribe a unique decomposition
of $\vec A$ into $\vec A_{\bot}$ and $\vec A_{\parallel}$. Under a gauge transformation of $\vec A$:
\begin{equation}
\vec A\longrightarrow \vec A^{\prime}=\vec A+\vec{\nabla} \chi,
\end{equation}
where $\chi$ is a nonsingular function,
$\vec A_{\bot}$ and $\vec A_{\parallel}$ transform as
\begin{eqnarray}
\vec A_{\bot}&\longrightarrow& \vec A^{\prime}_{\bot}=\vec A_{\bot}, \label{gaugetransAperp}\\
\vec A_{\parallel}&\longrightarrow& \vec A^{\prime}_{\parallel}=\vec
A_{\parallel}+\vec{\nabla} \chi. \label{gaugetransApara}
\end{eqnarray}
The condition $\vec{\nabla}\times \vec A_{\parallel}=0$ and Eq. (\ref{gaugetransApara}) tell us
that the longitudinal part $\vec A_{\parallel}$ is a pure gauge part
and it transforms in the same manner as does the full $\vec A$.
The transverse part $\vec A_{\bot}$ is gauge invariant under all gauge transformations and should be regarded as the "physical" part of ${\vec A}$.

Now, let us turn to the discussion of the A-B effect. The A-B effect, which indicates the importance of vector potential in quantum mechanics, has been widely studied over 50
years. In the literature the A-B effect is usually ascribed to the
non-trivial topology of the region outside the infinitely long solenoid. Here we shall study this problem from the perspective of the gauge potential decomposition (\ref{decomposition}).
We describe the system using cylindrical polar coordinates with z-axis
along the symmetry axis of the infinitely long solenoid.
According to Helmholtz theorem, ${\vec A}_{\perp}$ at all points of the space (both outside the solenoid and inside the solenoid)
can be expressed as
\begin{eqnarray}\label{ABAperp}
{\vec A}_{\perp}(x)&=&{\vec \nabla}\times \frac{1}{4\pi}\int d^3 x'
\frac{{\vec \nabla}'\times {\vec A}(x')}{|{\vec x}-{\vec x}'|}
\nonumber \\
&=& {\vec \nabla}\times \frac{1}{4\pi}\int d^3 x' \frac{{\vec
B}(x')}{|{\vec x}-{\vec x}'|} \nonumber \\
&=& {\vec \nabla}\times \frac{B}{4\pi} \vec{e}_z \int\limits_{\rho < R} \frac{d^3 x'}{|{\vec x}-{\vec x}'|},
\end{eqnarray}
where $R$ is the radius of the cross section of the solenoid. In the cylindrical polar coordinates, the field point $x=(r\cos\varphi,r\sin\varphi,z)$
and the source point $x'=(\rho\cos\varphi',\rho\sin\varphi',h)$.
The integral in Eq. (\ref{ABAperp}) can be written as
\begin{eqnarray}\label{ABintegral}
&&\int\limits_{\rho < R} \frac{d^3 x'}{|{\vec x}-{\vec x}'|}\nonumber\\
&=& \int\limits_{0}^{R} \rho d \rho \int\limits_{0}^{2\pi} d\varphi'  \int\limits_{-\infty}^{+\infty} dh
\frac{1}{\sqrt{r^2+\rho^2 -2 r \rho \cos(\varphi'-\varphi)+(h-z)^2}} \nonumber \\
&=& \int\limits_{0}^{R} \rho d \rho \int\limits_{0}^{2\pi} d\varphi'  \int\limits_{-\infty}^{+\infty} dh
\frac{1}{\sqrt{r^2+\rho^2 -2 r \rho \cos \varphi'+h^2}}.
\end{eqnarray}
For an infinitely long solenoid, the integral over $h$ in Eq. (\ref{ABintegral}) is logarithmically divergent. It can be regulated by introducing a large cutoff $L$:
\begin{eqnarray}
&& \int\limits_{-\infty}^{+\infty} dh
\frac{1}{\sqrt{r^2+\rho^2 -2 r \rho \cos \varphi'+h^2}} \nonumber \\
&\rightarrow& \int\limits_{-L}^{+L} dh
\frac{1}{\sqrt{r^2+\rho^2 -2 r \rho \cos \varphi'+h^2}} \nonumber \\
&=& \ln 4L^2 -\ln (r^2+\rho^2-2 r \rho \cos \varphi')+O(\frac{1}{L^2}).
\end{eqnarray}
Then
\begin{eqnarray}\label{ABintegral2}
&&\int\limits_{\rho < R} \frac{d^3 x'}{|{\vec x}-{\vec x}'|}\\
&&=\pi R^2 \ln 4L^2-\int\limits_{0}^{R} \rho d \rho \int\limits_{0}^{2\pi} d\varphi'
\ln (r^2+\rho^2-2 r \rho \cos \varphi')+O(\frac{1}{L^2}).\nonumber
\end{eqnarray}
The integral in the right-hand-side of Eq. (\ref{ABintegral2}) can be evaluated to be
\begin{eqnarray}
&&\int\limits_{0}^{R} \rho d \rho \int\limits_{0}^{2\pi} d\varphi'
\ln (r^2+\rho^2-2 r \rho \cos \varphi')\nonumber\\
&=& 2\pi \int\limits_{0}^{R} \rho d \rho \ln \frac{r^2+\rho^2 +|r^2-\rho^2|}{2} \nonumber \\
&=& \left\{\begin{array}{ll}
& \pi R^2 \ln R^2-\pi R^2+\pi r^2 ~~~~~r \leq R \\
& \pi R^2 \ln r^2  ~~~~~~~~~~~~~~~~~~~~~~~r>R
\end{array}
\right.
\end{eqnarray}
The divergent term $\pi R^2 \ln 4L^2$ in Eq. (\ref{ABintegral2}) is a constant and does not contribute to ${\vec A}_{\perp}(x)$. The contribution of the third term in Eq. (\ref{ABintegral2}) vanishes in the limit $L \rightarrow \infty$. We then obtain
\begin{equation}\label{Aperp}
{\vec A}_{\perp}(x) = \left\{
\begin{array}{ll}
& \frac{Br}{2} \vec{e}_\varphi ~~~~~~r \leq R \\
& \frac{\Phi}{2\pi r} \vec{e}_\varphi ~~~~~~r >R,
\end{array}
\right.
\end{equation}
where $\Phi=\pi R^2 B$ is the magnetic flux inside the solenoid, and $\vec{e}_\varphi$ is the base vector along the $\varphi$ direction of the cylindrical polar coordinate system. Note
that outside the solenoid ($r>R$) one has ${\vec A}_{\perp}(x)=\frac{\Phi}{2\pi}\vec{\nabla}\varphi$.
As is shown by Eq. (\ref{ABAperp}), once the magnetic field ${\vec B}$
at all points inside the solenoid has been measured, one can
determine ${\vec A}_{\perp}$ at all points of the space. In this
sense, ${\vec A}_{\perp}$ in A-B effect is completely determined by
the magnetic field $\vec{B}$ and is thus physical. It is not a pure gauge
term and cannot be gauged away by a gauge transformation. Here, we note that the transverse vector potential in electrodynamics was also studied in Ref. \cite{Dubovik}.
In that reference the authors argued that the transverse vector potential
is the physical degrees of freedom of electrodynamics and also discussed the transverse vector potential in the region where the field strength vanishes (such as the region outside the solenoid
in the case of A-B experiment). Compared with their approach, we do not discuss $\vec{A}_\perp$ inside the solenoid and outside the solenoid separately. As is shown above, we derive an explicit expression of $\vec{A}_\perp$ in the A-B experiment at all points (both inside the solenoid and outside the solenoid) by means of Helmholtz theorem. Our approach shows that the $\vec{A}_\perp$ inside the solenoid and outside the solenoid is a unified physical quantity and is the physical degree of freedom of the electromagnetic field. We also note that in Ref. \cite{Majumdar} the authors proposed a reformulation of electrodynamics in terms of a physical vector potential entirely
free of gauge ambiguities. Our formulation differs from theirs in that in our formulation the gauge degrees of freedom still exists, whereas in their formulation there are no gauge degrees of freedom at all. 

Recently, based on the gauge potential decomposition (\ref{decomposition}), the authors of \cite{Wang} proposed that the proper quantum mechanical momentum operator for a charged particle
in an external magnetic field should be $\vec{P}={\vec p}-q{\vec
A}_{\parallel}$ and the corresponding orbital angular momentum
operator is $\vec{L}=\vec{r}\times({\vec p}-q{\vec A}_{\parallel})$.
Thus, the proper momentum operator for the electron in A-B experiment is
\begin{eqnarray}\label{momentum}
\vec{P}&=&{\vec p}-q {\vec A}_{\parallel}={\vec p}-q{\vec A}+q{\vec A}_{\bot} \nonumber \\
&=& m{\vec v}+\frac{q\Phi}{2\pi r}\vec{e}_\varphi,
\end{eqnarray}
where $m{\vec v}={\vec p}-q{\vec A}$ is the mechanical momentum. Note that
$m{\vec v}$ is observable only in classical physics. In quantum mechanics,
when the magnetic field $\vec{B}$ is nonzero, the three components of $m{\vec v}$ do not commute with each other and therefore cannot be measured simultaneously.
Here we also point out that neither the mechanical momentum nor the canonical momentum provides for the true description of the motion of an electron wave passing around an infinite magnetic solenoid,
whereas the proper momentum operator given in Eq. (\ref{momentum}) does precisely what is needed. It displays that an electron wave passing on one side of the solenoid picks up additional momentum
while the electron wave passing on the other side loses momentum. Thus fringes appear when the waves are recombined on the far side. Therefore, A-B effect is due to interaction with ${\vec A}_\perp$,
which is shown clearly by the expression (\ref{momentum}) for the proper momentum operator.

On the other hand, in the discussion of A-B experiment a multi-valued vector potential can
also be defined \cite{a5}
\begin{equation}\label{multivaluedA}
     \vec A'=-\varphi B_z(r)\vec{r},
\end{equation}
where $B_z(r)=B \theta(R-r)$ is the z-component of the magnetic field.
The vector potential in Eq. (\ref{multivaluedA}) also satisfies
\begin{equation}
     \vec{\nabla} \times \vec A'=B_z(r)\vec{e}_z=\vec{B}.
\end{equation}
It can be easily seen that $\vec A'$ differs from $\vec A_{\bot}$ by
a pure gauge term:
\begin{equation}
     \vec A'= {\vec A}_{\perp}-\vec{\nabla}(\frac{\Lambda(r) \varphi}{2\pi}),
\end{equation}
where
\begin{equation}
     \Lambda(r)=2\pi\int_{0}^{r}r'B_z(r')dr'.
\end{equation}
It is obvious that only the purely transverse part $\vec
A_{\bot}$ of the gauge potential $\vec A'$ is
physical. In fact, there are various other expressions of the vector potential that are used to describe the A-B effect.
All of them are connected to $\vec{A}_\perp$ in Eq. (\ref{Aperp}) by a suitable gauge transformation.

Now we turn to the discussion of A-H mechanism in G-L model for superconductors.
We begin with the Lagrangian for the relativistic version of G-L model
\begin{equation}\label{Lagrangian}
{\cal L}=-\frac{1}{4}F_{\mu \nu }F^{\mu \nu }+(D^\mu \phi)^{\ast} (D_{\mu
}\phi) -V(\phi),
\end{equation}
where the complex scalar field $\phi$ is the order parameter and
$D_{\mu}= \partial _{\mu }+iq A_{\mu }$ is the covariant
derivative with $q=-2e$ being the charge of the Cooper pair. Here the potential $V(\phi)=\alpha \left| \phi \right|
^{2}+\frac{\beta }{2}\left| \phi \right|^{4}$. According to Landau phase transition
theory, the coefficient $\beta$ is always positive. For $\alpha >0$ the potential has a minimum at $|\phi|=0$ and the system is in the symmetric phase. For $\alpha <0$ the potential has a minimum at $|\phi|^2=-\frac{\alpha}{\beta}$. In this case, the vacuum is degenerate
and spontaneous symmetry breaking occurs.

In the case of symmetry breaking, $\phi$ acquires a nonzero vacuum value:
\begin{equation}
\left\langle \phi \right\rangle =\phi _{0}=\sqrt{-\frac{\alpha
}{\beta }}.
\end{equation}
The scalar field can be written as
\begin{equation}\label{phi}
\phi \left( x\right) =\left( \phi _{0}+\eta \left( x\right) \right)
e^{i\theta \left( x\right) },
\end{equation}
where $\eta \left( x\right) $ and $\theta \left( x\right)$ are the
amplitude and phase fluctuations of the order parameter,
respectively, and the latter represents the massless Goldstone mode.
When expressed in terms of $\eta$ and $\theta$, the Lagrangian (\ref{Lagrangian}) reads
\begin{eqnarray}\label{Lagrangian2}
{\cal L}&&=-\frac{1}{4}F_{\mu \nu }F^{\mu \nu }+
\partial _{\mu}\eta \partial^{\mu }\eta-V(\phi_{0}+\eta)\nonumber\\
&&+(\phi_{0}+\eta)^{2}(\partial_{\mu}\theta -2eA_{\mu})(\partial^{\mu}\theta
-2eA^{\mu}).
\end{eqnarray}
One can further absorb the Goldstone field $\theta$ into $A_\mu$ by defining a new field $A_\mu'$:
\begin{equation}
A_\mu'=A_\mu-\frac{1}{2e}\partial_\mu \theta.
\end{equation}
Then the Lagrangian (\ref{Lagrangian2}) reads
\begin{equation}\label{Lagrangian3}
{\cal L}=-\frac{1}{4}F'_{\mu\nu}F'^{\mu\nu} +
\partial_{\mu}\eta \partial^{\mu}\eta -V(\phi_{0}+\eta)
 +4e^{2}(\phi_{0}+\eta)^{2}A'_{\mu}A'^{\mu},
\end{equation}
where $F'_{\mu\nu}=\partial_\mu A'_\nu-\partial_\nu A'_\mu$. In the Lagrangian (\ref{Lagrangian3}), the Goldstone field does not appear and
the original massless gauge field acquires a mass. The degree of freedom of the Goldstone field has transformed into the longitudinal component of the massive gauge field.
This is just the A-H mechanism.

Now we shall discuss the A-H mechanism from the perspective of the gauge potential decomposition Eq.(\ref{decomposition}). It is well known that the G-L order parameter for
superconductors is represented by the bosonic field $\phi(\vec x)=\sqrt{\rho(\vec x)}e^{i\theta(\vec x)}$, where the amplitude
$\rho(\vec x)$ is meaningful and observable. It can be identified as
the density of Cooper pairs. In the following we shall argue that for a bulk superconductor, the gradient of the phase field $\theta$ provides the longitudinal
component ${\vec A}_{\parallel}$ of the vector potential ${\vec A}$, while for a superconducting ring,
$\theta$ can be decomposed into a nonsingular longitudinal field $\theta_{1}$
and a singular transverse field $\theta_{2}$, the gradient of $\theta_{1}$ providing the longitudinal
component ${\vec A}_{\parallel}$, and $\theta_2$ being connected with the transverse component $\vec{A}_{\perp}$ which is induced by a constant magnetic flux $\Phi$ trapped in the
superconducting ring.

From the G-L wave function, we can first obtain the electric current density inside the superconductor
\begin{equation}\label{current}
 \vec j=\frac{q\rho}{m}(\hbar\vec{\nabla}\theta-\frac{q}{c}\vec A),
\end{equation}
where $\hbar\vec{\nabla}\theta=m{\vec V}$ is the usual
canonical momentum for the Cooper pair.  We now rewrite the current
density as
\begin{equation}
\vec j=\frac{q\rho}{m}(\hbar\vec{\nabla}\theta-\frac{q}{c}{\vec A}_{\parallel}-\frac{q}{c}{\vec A}_{\perp}).
\end{equation}
Since $\vec j$ and $\vec{A}_\perp$ are gauge invariant quantities, whereas $\vec{\nabla}\theta$ and ${\vec A}_{\parallel}$ are not, so for a bulk superconductor one should have the relation
\begin{equation}\label{relation}
\hbar\vec{\nabla}\theta=\frac{q}{c}{\vec A}_{\parallel}.
\end{equation}
Eq. (\ref{relation}) shows clearly that in a bulk superconductor the gradient of the phase field provides the longitudinal
component of the vector potential. In this sense, this relation just plays the same role as the A-H mechanism does in the previous discussion, i.e., the Goldstone field $\theta$ is eaten up and the longitudinal component
${\vec A}_\parallel$ appears.
Note that owing to Eq. (\ref{relation}) we do not need to adopt Coulomb gauge to eliminate the pure gauge term ${\vec A}_{\parallel}$ in G-L theory.
Eq. (\ref{relation}) immediately leads to the famous London equation
\begin{equation}\label{London}
\vec j=-\frac{q^{2}\rho}{mc}\vec A_{\bot}.
\end{equation}
Substituting London equation into the Maxwell equation (for static fields)
\begin{equation}\label{Maxwell}
\vec{\nabla}\times \vec{B}=\frac{4\pi}{c}\vec{j},
\end{equation}
and using $\vec{\nabla}\times (\vec{\nabla} \times  \vec{A}_\perp)=-\nabla^2  \vec{A}_\perp$, we can obtain
\begin{equation}\label{Meissner}
\nabla^2 \vec{A}_\perp-\frac{1}{\lambda^2}\vec{A}_\perp=0,
\end{equation}
where $\lambda=(\frac{mc^2}{4\pi q^2 \rho})^{1/2}$.
Eq. (\ref{Meissner}) leads to the exponentially decaying behavior of the magnetic field in a bulk superconductor and $\lambda$ is the penetration depth. Therefore, inside a bulk superconductor
$\vec A_{\bot}=0$, $\vec B=0 $, and $\vec j=0$.
This is just the Meissner effect.

Now let us apply the G-L theory to the case of a multiply connected superconducting
ring. We describe the system using cylindrical polar coordinates
with z-axis perpendicular to the plane of the ring. We assume that
there is a magnetic flux $\Phi$ through the ring. Below $T_{c}$ a
persistent current will flow around the ring to maintain the
constant flux $\Phi=n\Phi_{0}$ in the ring, where $\Phi_{0}=\frac{hc}{2e}$ is
the flux quantum. Due to the A-B effect $\vec A_{\bot}\neq0$ inside the
superconducting ring, which differs from the case of a bulk
superconductor.

For a superconducting ring the phase field can be decomposed into two parts:
$\theta=\theta_{1}+\theta_{2}$, where the longitudinal phase field
$\theta_{1}$ is nonsingular and the transverse phase field
$\theta_{2}$ is singular. $\theta_{1} $ and $\theta_{2}$ are
independent of each other. Here the nonsingular phase field $\theta_{1}$
still represents the Goldstone mode and the singular phase field
$\theta_{2}\neq0$ is multi-valued \cite{Liu}. Such a decomposition of
phase field can also be seen in Kosterlitz-Thouless (K-T) transition
\cite{a6}. In that case, $\theta_{1}$ is the analytic spin-wave component and
$\theta_{2}$ represents the singular vortex component. Note that
K-T transition only occurs in 2D space.
For the case of a superconducting ring, similar to Eq. (\ref{relation}), we should have
\begin{equation}\label{relationring}
\hbar\vec{\nabla}\theta_{1}=\frac{q}{c}\vec A_{\parallel},
\end{equation}
Eq. (\ref{relationring}) leads to
\begin{equation}\label{ringcurrent}
\vec j=\frac{q\rho}{m}(\hbar\vec{\nabla}\theta_{2}-\frac{q}{c}\vec
A_{\perp}).
\end{equation}
Substituting Eq. (\ref{ringcurrent}) into Eq. (\ref{Maxwell}), one can obtain
\begin{equation}\label{Meissnerring}
\nabla^2 \vec{A}_{\perp}-\frac{1}{\lambda^2}\vec{A}_{\perp}=\frac{\Phi_0}{2\pi \lambda^2} \vec{\nabla}\theta_2.
\end{equation}
The solution to Eq. (\ref{Meissnerring}) can be written as
$\vec A_{\perp}=\vec A_{1 \perp}+\vec A_{2 \perp}$, where $\vec A_{1
\perp}$ is a solution to the homogeneous equation corresponding to
Eq. (\ref{Meissnerring}) and $\vec A_{2 \perp}$ is a specific
solution to Eq. (\ref{Meissnerring}). For the 1D case it is easy to obtain
$\vec A_{1 \perp}=\vec{A}_{1 \perp}(0)e^{-\frac{x}{\lambda}}$. This is
just the Meissner effect. The specific solution $\vec A_{2
\perp}$ is connected with the A-B effect. If one takes $\theta_2=n \varphi$, where $n$ is an integer (this is due to the single-valuedness condition of the wavefunction), then it can be verified that a specific
solution to Eq. (\ref{Meissnerring}) is $\vec A_{2 \perp}=-\frac{n \Phi_0}{2\pi}\vec{\nabla}\varphi$ (inside the superconducting ring). Here we have used the equality $\nabla^2 \varphi=0$. It is obvious that $\vec A_{2 \perp}$ is a solution for
A-B effect with $-n \Phi_0$ being the flux trapped in the superconducting ring. It gives null magnetic field inside the superconducting ring. The complete solution in this case is the superposition of the Meissner effect solution and A-B effect solution. Therefore, inside a
superconducting ring $\vec B=0$ and $\vec A_{1 \perp}=0$, whereas
$\vec A_{2 \perp}\neq0$. Note that here we have given a unified description of Meissner effect and A-B effect for the case of a superconducting ring.

Owing to Maxwell equation (\ref{Maxwell}),
inside the superconducting ring $\vec{j}=0$.
From Eq. (\ref{ringcurrent}) it follows immediately that inside a superconducting ring 
\begin{equation}\label{ring}
\hbar\vec{\nabla}\theta_{2}=\frac{q}{c}\vec A_{\bot}.
\end{equation}
Taking the line integral of both sides of Eq. (\ref{ring}) along a curve $\Gamma$ around the
superconducting ring, one obtains 
\begin{equation}
\oint_{\Gamma}\vec{\nabla}\theta_{2} \cdot d {\vec l}=\oint_{\Gamma}\frac{q}{\hbar c}\vec A_{\bot}\cdot d\vec{l}.
\end{equation}
If one assumes that $\theta_1$ is single valued, then from this one immediately obtains the magnetic flux quantization
condition $\Phi=n\Phi_{0}$ by the wave function single-valuedness
condition $\phi( r,\varphi,z)=\phi(r,\varphi+2\pi,z)$. Note that in Ref. \cite{Dubovik} the authors also gave a discussion on the wave function of the ground state of a superconducting ring, while in our work we have given a unified description of Meissner effect and flux quantization for the case of
a superconducting ring from the perspective of gauge potential decomposition.

To summarize, in this paper we study the A-B effect and A-H
mechanism in G-L model of superconductors from the perspective of
the decomposition of U(1) gauge potential. By the Helmholtz theorem,
we derive exactly the expression of the transverse gauge potential
$\vec{A}_\perp$ in A-B experiment, which is gauge-invariant and
physical. For the case of a bulk superconductor, we find that the
gradient of the total phase field $\theta$ of the order parameter
provides the longitudinal component ${\vec A}_{\parallel}$ of the
vector potential, which reflects the A-H mechanism, while the transverse part $\vec A_{\bot}$ only exists in the surface region, which reflects the Meissner effect. For the case of
a superconducting ring, the phase field can be decomposed into two
parts: $\theta=\theta_{1}+\theta_{2}$, where the longitudinal phase
field $\theta_{1}$ is nonsingular and the transverse phase field
$\theta_{2}$ is multi-valued and singular. The gradient of the phase
field $\theta_1$ provides the longitudinal component ${\vec
A}_{\parallel}$, while $\theta_2$ will produce new physical effects,
for example, the flux quantization inside a superconducting ring.

\bigskip

\textbf{Acknowledgement:}

\bigskip
This work is supported in part by the
National Science Foundation of China (under Grant No.11075075 and
10935001), the Research Fund for the Doctoral Program of Higher
Education (under Grant No.200802840009) and a project funded by the Priority
Academic Program Development of Jiangsu Higher Education Institution.


\begin{references}
\bibitem{Duan} Y.S. Duan and M.L. Ge, Sci. Sin. {\bf 11}, 1072(1979).
\bibitem{Cho} Y.M. Cho, Phys. Rev. Lett. {\bf 46}, 302 (1981).
\bibitem{Faddeev} L. Faddeev and A.J. Niemi, Phys. Rev. Lett. {\bf 82}, 1624 (1999).
\bibitem{a4} X.S. Chen, X.F. L\"{u}, W.M. Sun ,F. Wang and T. Goldman, Phys. Rev. Lett. {\bf100}, 232002(2008).
\bibitem{Wang} F. Wang, X.S. Chen, X.F. L\"{u}, W.M. Sun and T. Goldman, arXiv: 0909.0798 [hep-ph].
\bibitem{Sun} W.M. Sun, X.S. Chen, X.F. L\"{u} and F. Wang, Phys. Rev. {\bf A82}, 012107 (2010).
\bibitem{a1} Y. Aharonov and D. Bohm, Phys. Rev. {\bf115}, 485 (1959).
\bibitem{a2} P.W. Anderson, Phys. Rev. {\bf130}, 439 (1963); P.W. Higgs, Phys. Lett. {\bf 12}, 132(1964).
\bibitem{a3} V.L. Ginzburg and L.D. Landau, Sov. Phys. JETP. {\bf20}, 1064 (1950).
\bibitem{Dubovik} V.M. Dubovik and S.V. Shabanov, J. Phys.{ \bf A23}, 3245 (1990).
\bibitem{Majumdar} P. Majumdar and S. Bhattacharjee, arxiv: 0903.4340 [hep-th].
\bibitem{a5} M. Bawin and A. Burnel, J. Phys. {\bf A16},2173(1983).
\bibitem{Liu}G.-Z. Liu and G. Cheng, Phys. Rev. {\bf B65}, 132513 (2002).
\bibitem{a6} J.M. Kosterlitz and D.J. Thouless, J.Phys.{\bf C662}, 1181 (1973).


\end{references}
\end{document}